\documentclass[sigconf]{acmart}

\usepackage{booktabs} 

\setcopyright{rightsretained}

\usepackage{enumitem}
\usepackage[linesnumbered,ruled]{algorithm2e}

\usepackage{multirow}
\usepackage{hhline}
\usepackage{mathtools}
\usepackage{amsmath}
\usepackage{color}
\usepackage{epsfig}
\usepackage{epstopdf}
\usepackage{dblfloatfix} 
\usepackage{graphicx}

\begin{document}

\copyrightyear{2018} 
\acmYear{2018} 
\setcopyright{acmcopyright}
\acmPrice{15.00}
\acmDOI{10.1145/3195970.3196001}
\acmISBN{978-1-4503-5700-5/18/06}

\title{BLASYS: Approximate Logic Synthesis Using\\ Boolean Matrix Factorization}

\author{Soheil Hashemi}
\affiliation{
  \institution{School of Engineering\\ Brown University}
  \city{Providence} 
  \state{RI 02912}
}
\email{soheil\_hashemi@brown.edu}

\author{Hokchhay Tann}
\affiliation{
  \institution{School of Engineering\\ Brown University}
  \city{Providence} 
  \state{RI 02912} 
}
\email{hokchhay\_tann@brown.edu}

\author{Sherief Reda}
\affiliation{
  \institution{School of Engineering\\ Brown University}
  \city{Providence} 
  \state{RI 02912} 
}
\email{sherief\_reda@brown.edu}

\begin{abstract}
Approximate computing is an emerging  paradigm where design accuracy can be traded off for benefits in design metrics such as design area, power consumption or circuit complexity. In this work, we present a novel paradigm to synthesize approximate circuits using Boolean matrix factorization (BMF). In our methodology the truth table of a sub-circuit of the design is approximated using BMF to a controllable approximation degree, and the results of the factorization are used to synthesize a less complex subcircuit. To scale our technique to large circuits, we devise a circuit decomposition method and a subcircuit design-space exploration technique to identify the best order for subcircuit approximations. Our method leads to a smooth trade-off between accuracy and full circuit complexity as measured by design area and power consumption. Using an industrial strength design flow, we extensively evaluate our methodology on a number of testcases, where we demonstrate that the proposed methodology can achieve up to 63\% in power savings, while introducing an average relative error of 5\%. We also compare our work to previous works in Boolean circuit synthesis and demonstrate significant improvements in design metrics for same accuracy targets.

\end{abstract}

\maketitle

\section{Introduction}
\label{sec:introduction}
Approximate computing is an emerging paradigm that trades off accuracy with improvements in power consumption, hardware complexity, or design area. Approximate computing is effective in applications that have inherent resilience to errors, such as signal processing, machine learning, computer vision, and computer graphics. As data-rich applications continue to rise, the relevance and need for approximate computing will increase. 

A key problem in approximate computing is how to generate or synthesize an approximate circuit given as inputs an existing circuit, presumably accurate. 
There are two lines of research in approximate synthesis. The first line of research has devised custom approximate designs for typical arithmetic building blocks (e.g., adders, multipliers \cite{Kahng,drum,Rehman,Imani1}). The second line has targeted approximation of more generic circuits either from gate-level (i.e., Boolean descriptions) \cite{LiL15,salsa,aslan14,sasimi,Miao13}, higher-level descriptions, such as RTL or behavioral descriptions \cite{Nepal14}, or even direct C to approximate hardware synthesis \cite{Lee17}.

This paper seeks to devise a new direction for approximate boolean-level circuit synthesis. Our inspiration comes from Boolean Matrix Factorization (BMF) that factors a boolean matrix into two boolean matrices~\cite{Miettinen11,Miettinen14}. BMF is a derivative of non-negative matrix factorization (NNMF), in which the elements of all input and output matrices are limited to the non-negative space~\cite{Lee99}. The non-negativity constraints on the factorization arise in physical domains, such as computer vision and document clustering~\cite{XuL03}. Recent advances in the mathematical community extends NNMF techniques to Boolean matrices, where matrix operations are carried in $GF(2)$~\cite{Miettinen11,Miettinen14}. The use of BMF as a technique for logic synthesis is a new direction in the field, and we show that it provides a solid foundation for approximate logic synthesis. We summarize our contributions as follows.

\begin{itemize}[leftmargin=5mm]
\item We devise a new  methodology for  {\bf B}MF-based {\bf L}ogic {\bf A}pproxi-mate {\bf SY}nthesi{\bf S} (BLASYS) that is  based on solid mathematical foundations, where Boolean Matrix Factorization (BMF) is used to generate approximate circuits with controllable trade-off between accuracy and circuit complexity.  
\item We modify existing BMF algorithms to incorporate the ability to work with different quality-of-results (QoR) functions, instead of the standard $L_2$ norm.
\item To scale the factorization method to large circuits, we propose a circuit decomposition method to break down a given circuit into manageable subcircuits with limited number of inputs and outputs. We propose a design-space exploration heuristic to order the subcircuits to identify a good sequence for generating their approximate variants. Our technique results in a very smooth trade-off between accuracy and circuit complexity.
\item We implement our approach and test it on a number of application circuits that are typically used in approximate computing domains. We show that our approach is able to trade-off accuracy with circuit area and power consumption as evaluated by an industry-strength synthesis tool. We also evaluate our methodology against a established approach in the literature (e.g., SALSA~\cite{salsa}) and show significant improvements.  
\end{itemize}

The organization of this paper is as follows. First, we overview related work in Section \ref{sec:prev_work}. We discuss the details of our proposed method in Section \ref{sec:methodology}, where we describe the basic idea of using BMF algorithms to approximate logic circuits, and then show how to scale our proposed method to larger circuits. We provide comprehensive results of our method's performance together with a comparison against a previous technique in Section~\ref{sec:results}. Finally, we summarize our main conclusions and directions for future work in Section~\ref{sec:conclusions}.

\section{Previous Work}
\label{sec:prev_work}
Recent work on approximate circuit synthesis can be divided into two general categories: Boolean or gate-level approaches and high-level synthesis approaches.

For Boolean and gate-level synthesis, a number of approaches have been proposed \cite{salsa,aslan14,sasimi,Miao13,Vasicek2016}. In SALSA, a systematic approach for approximate circuit synthesis is proposed \cite{salsa}. The idea is to create a difference circuit that compares the QoR between the original circuit and the approximated circuit. The don't cares of the outputs of the approximate circuit -- which are internal nodes in the difference circuit -- with respect to outputs of the difference circuit can be used to simplify the approximate circuit using regular logic synthesis techniques. This approach has been extended in ASLAN~\cite{aslan14} to model error arising over multiple cycles. ASLAN also uses a circuit block exploration method that identifies the impact of approximating the combinational blocks and then uses a gradient-descent approach to find good approximations for the entire circuit. In SASIMI ~\cite{sasimi}, a technique is proposed to identify similar signals, such that their values agree over a large number of input test cases, and then substitute one for the other, simplifying the logic. A logic synthesis formulation proposed by Miao {\it et al.} uses a two-level logic synthesis approach that incorporates constraints on error deviation, and then a heuristic is used to solve the synthesis formulation \cite{Miao13}. Evolutionary techniques have been also explored~\cite{Vasicek2016}.

For high-level logic synthesis, ABACUS seeks to generate variants of an input high-level Verilog description file by applying a set of possible transformations, such as  bit width truncation, operand simplification and variable-to-constant substitution, to generate a set of mutant approximate circuit variants \cite{Nepal14}. A multi-objective design space exploration technique is used to identify the best set of approximate variants.  Recently, a new technique is proposed to raise the level of abstraction by synthesizing approximate circuit directly from C descriptions \cite{Lee17}. High-level synthesis in conjunction with approximations on the critical path can yield additional savings through voltage scaling \cite{TETC16,Lee17}.

\section{Proposed Methodology}
\label{sec:methodology}
\begin{figure}[t!]
\vspace{-0.1in}
	\begin{center}
		\includegraphics[scale=0.18]{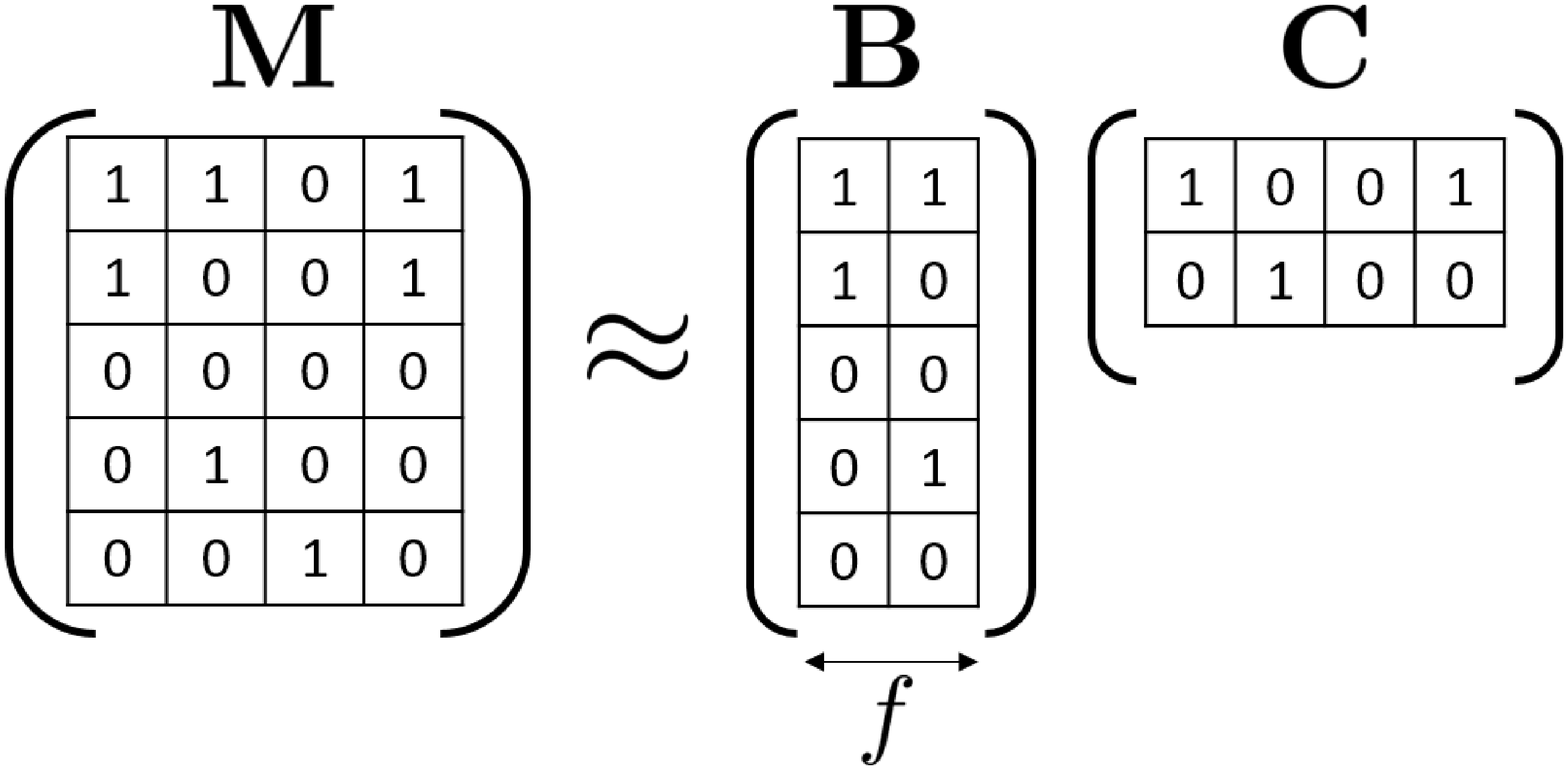}
		\vspace{-0.1in}
		\caption{Boolean NNMF example.}
		\label{fig:nmf}
	\end{center}
    \vspace{-0.1in}
\end{figure}

Non-negative matrix factorization (NNMF) is a factorization technique where a $k\times m$ matrix $\mathbf{M}$ is factored into two non-negative matrices: a $k\times f$ matrix $\mathbf{B}$, and an $f \times m$ matrix $\mathbf{C}$, such that $\mathbf{M} \approx \mathbf{B} \mathbf{C}$ \cite{Lee99}. The non-negativity constraints on the factorization enables the utilization of the factorization algorithm in physical domains, such as computer vision and document clustering. NNMF essentially compresses the data storage in an approximate way depending on the {\it factorization degree} ($f$)~\cite{XuL03}. In the mathematical statistics community, the {\it factorization degree} determines the number of ``features" that are computed~\cite{Miettinen11}. Therefore, $f$, clearly represents a trade-off between quality of factorization and storage amount. 
Recently, NNMF has been extended to boolean matrices where elements of all matrices are restricted to Boolean values. In this case, multiplications can be performed using logical AND, and additions are performed using logical OR (for Boolean semi-ring implementations) and logical XOR (for Boolean field implementations) \cite{Miettinen11,Miettinen14}.
Figure \ref{fig:nmf} provides an example of NNMF over GF(2).

\begin{figure}[b!]
	\begin{center}
		\includegraphics[scale=0.18]{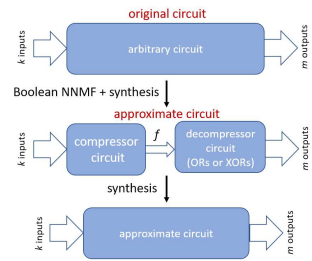}
		\caption{Generating approximate circuits using BMF.}
		\label{fig:bmf}
	\end{center}
    \vspace{-0.1in}
\end{figure}

\begin{figure*}[t!]
	\begin{center}
		\includegraphics[scale=0.48]{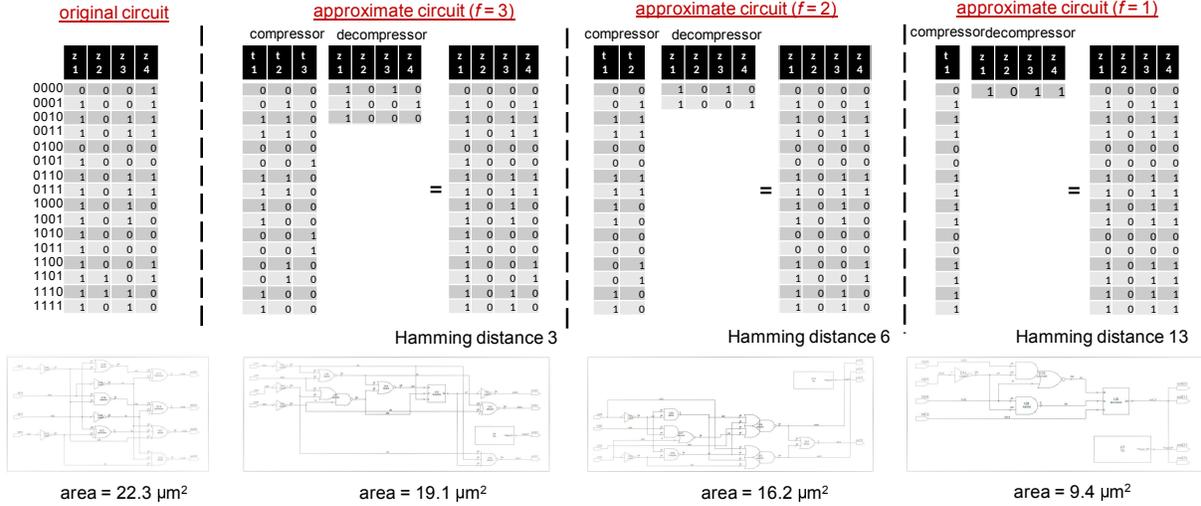}
		\vspace{-0.1in}
		\caption{Results of proposed approximation method with various $f$ on a simple circuit  for illustration purposes. Circuits are synthesized using Synopsys DC using 65 nm technology library. A semi-ring implementation is used for Boolean NNMF.}
		\label{fig:bmf_illust}
	\end{center}
		\vspace{-0.1in}
\end{figure*}

\subsection{Circuit Approximation using BMF}

In our proposed approach, a multi-output logic circuit with $k$ inputs and $m$ outputs is first evaluated to generate its truth table. The truth table, represented by $\mathbf{M}$, is then given as input to a BMF algorithm together with the target factorization degree $1\leq f < m$, to produce the two factor matrices $\mathbf{B}$ and $\mathbf{C}$. Matrix $\mathbf{B}$ is then given as the input truth table to a logic synthesis tool to generate a $k$ input, $f$ output circuit, which we refer to as the {\it compressor circuit}. Note that the compressor matrix is simply the truth table of a circuit with the same number of inputs as the original circuit but with fewer ($f$ to be exact) output signals. Therefore, it can easily be mapped to logic. These $f$ outputs from the compressor circuit are then combined by the {\it decompressor circuit} according to the $\mathbf{C}$ matrix using a network of OR gates (for Boolean semi-ring implementations) or XOR gates (for Boolean field implementations), to generate the approximate $m$ outputs. More specifically, a $1$ in the $(i,j)$ index of the decompressor matrix represents the existence of the $f_i$ intermediate signal in the $j$-th output, effectively mapping each one to a OR (or XOR) gate. Using this methodology, any arbitrary circuit can be approximated by forcing the circuit to compress as much information as possible in $f$ intermediate signals and then decompress them using simple OR (or XOR) gates. Figure~\ref{fig:bmf} illustrates the proposed approach. 

Figure \ref{fig:bmf_illust} provides an illustrative example of a 4-input, 4-output arbitrary logic circuit. First, we present the original circuit with its truth table, and we synthesize it with Synopsys Design Compiler (DC) using 65 nm library. We then provide approximate variants for the circuit with $f=1$, $f=2$, and $f=3$. We computed the truth tables for the compressor and decompressor using the ASSO NNMF algorithm~\cite{Miettinen11,Miettinen14}. We provide both the quality of results as measured by the Hamming distance between the truth table of the original circuit and the approximate circuit as well as the design area reported by DC. For instance, when $f=3$, we reduce the area of the circuit by 14.3\%, while compromising the quality of results (QoR) by only 4.6\% since the Hamming distance between the original and the approximate truth tables is equal to 3; that is out of the 64 entries in the truth table, only 3 entries flipped in the approximate circuit. With $f=2$ and $f=1$, we can reduce the area by 27.3\% and 57.8\% while compromising the QoR by 9.3\% and 20.3\% respectively.

Our approach leads to a new paradigm for creating approximate logic circuits in a controlled fashion that are based on solid mathematical foundations. There are two main challenges:

\begin{enumerate}
\item NNMF algorithms use the $L_2$ norm to measure the quality of factorization. For Boolean matrices, $L_2$ translates to Hamming distance. In addition, we need to identify methods to factorize for other QoR metrics that are relevant for approximate applications. 
\item The basic idea is limited in scalability since the complexity of generating the truth table grows exponentially as a function of the number of circuit's inputs. Thus, we need to create factorization methods that can scale up for large circuits.
\end{enumerate}

\subsection{Factorization for Arbitrary QoR}

The goal of the BMF algorithm is to minimize $||\mathbf{M}-\mathbf{BC}||_2$, which  translates to Hamming distance in $GF(2)$. However, not all applications or circuits necessarily use this metric to assess QoR. For instance, in the case of circuit design, if an $m$ bit signal is to be interpreted as an $m$ bit number, Hamming distance is not really an accurate representation of the inaccuracies as mismatches in different bit indices contribute differently to the actual error. 
 
To take into account the non-uniform nature of  bit significance, we propose to modify the  NNMF algorithms in the literature to account for bit indices. More specifically, instead of minimizing $||\mathbf{M}-\mathbf{B}\mathbf{C}||_2$, we propose minimizing $||(\mathbf{M}-\mathbf{B}\mathbf{C})\mathbf{w}||_2$, where $\mathbf{w}$ is a constant weight vector. For example, if numerical difference is the target QoR, then the $\mathbf{w}$ vector will be based on powers-of-two (e.g., 8, 4, 2, 1) therefore reflecting the fact that different bit positions lead to different numerical weights. In this work, we modify the ASSO~\cite{Miettinen14} algorithm as such to penalize mismatches on the higher bit locations more than on the less significant bits. In Section~\ref{sec:result1}, we demonstrate how such weighting scheme can improve the results compared to the uniformly weighted standard BMF algorithms.

\subsection{Scaling Up for Large Circuits}
\label{sec:large_cir}

Calculating the BMF is limited by computational complexity as one needs to generate the truth table for every possible input and state combination. Furthermore, BMF is a NP-hard problem, and most algorithms in the literature are heuristics \cite{Lee99,Miettinen11,Miettinen14}. We propose a simple approach to scale BMF calculations for larger circuits. The basic idea is to decompose a large circuit into a number of subcircuits each with a maximum of $k$ inputs and $m$ outputs as afforded by the runtime of the factorization algorithm and memory requirements. Note that this approach is reminiscent {\it but yet fundamentally different} than FPGA mapping algorithms, where the goal is to map a circuit into logic elements, each with limited number of inputs \cite{Cong94}. Our motivation for decomposition is different because (1) we are mapping to subcircuits purely to address computational complexity, and (2)  we apply the BMF on the truth tables of the subcircuits, and then we synthesize the factored circuits into {\it any} target ASIC or FPGA technology. Instead of using classical $k$-cut algorithms, e.g. \cite{Cong94}, we propose to use $k\times m$-cut algorithms (e.g., KL algorithm \cite{Martinello10}) to identify subcircuits with a maximum input of $k$ and maximum output of $m$. Note that $k$ and $m$ are design choices mostly determined by the runtime and memory budgets.

Decomposing a large circuit into subcircuits of size $k\times m$ requires changing the way we evaluate the QoR. In particular one cannot evaluate the QoR of a subcircuit in isolation from the rest of the circuit, since a small error in the output of the subcircuit can propagate leading to larger errors.  Thus, instead of evaluating the QoR of an original subcircuit against its approximate version,   we have to evaluate QoR of the entire circuit $Cir(s_i\rightarrow T_{s_i,f_i})$, where $Cir(s_i\rightarrow T_{s_i,f_i})$ represents the approximate circuit created by substituting an accurate subcircuit, $s_i$, with its approximate version, $T_{s_i,f_i}$, using a $f_i$ factorization degree. As evaluating the entire circuit for all possible inputs  is infeasible, we use Monte Carlo sampling to estimate the QoR of the approximate version of the entire circuit. 

\setlength{\textfloatsep}{4pt}
\begin{algorithm}[t!]
	\footnotesize
	\SetKwBlock{Begin}{begin}{end}
	\SetKwInOut{Input}{Input}
	\SetKwInOut{Output}{Output}
	\Input{Accurate Circuit $ACir$, Error Threshold}\label{alg:input}
	\Output{Approximate Circuit $Cir$}
	subcircuits=Decompose input circuit using $k\times m$ decomposition\\
    // Factorization profiling Phase\\
    \For{each subcircuit $s_i$ with $m_i \leq m$ outputs }
	{
		$\mathbf{M}$=Construct truth table of $s_i$\\
        // profile for every possible factorization degree\\
          \For{$f$=1 to $m_i$-1}
          {
            $[\mathbf{B}, \mathbf{C}]$ = BMF($\mathbf{M}$, $f$)\\
            $T_{s_i,f}$=Construct truth table of $\mathbf{B}\mathbf{C}$\\
    	  }
    }
    // Circuit Space Exploration Phase\\
    Cir=ACir;\\
    Let $f_i=m_i$ for all subcircuits $s_i$\\
    \While{$QoR(Cir)$ < threshold}
    {
	\For{each subcircuit $s_i$ with $f_i>1$}
	{
        $Cir'$=$Cir(s_i\rightarrow T_{s_i,f_i-1})$\\
        $\Delta err_i$ = $QoR(Cir')- QoR(Cir)$\\
    }
    $b =  \arg\min_{i} (\Delta err_i$) \\
    $Cir = Cir(s_b \rightarrow T_{s_b,f_b-1})$\\
    $f_b = f_b-1$\\
    }
    $Cir$=Synthesize Best new Design\\
	\Return $Cir$
	\caption{BLASYS: Boolean Level Approximate Circuit Synthesis}
		\label{alg:order}
\end{algorithm}

In addition, the order of processing the subcircuits and  the target factorization degree for each subcircuit is an important consideration. We devise Algorithm~\ref{alg:order} to gradually approximate the circuit as guided by circuit accuracy. After identifying the subcircuits (Line 1), the first stage of the algorithm (lines 3-10) calculates the {\it potential} approximate versions for each subcircuit under various factorization degrees. The next stage (lines 12-22) seeks to explore the space of potential approximate subcircuits to identify a good approximation order. Lines 15-18 assess the reduction in accuracy of the entire circuit if the degree of factorization of each subcircuit is decremented. The subcircuit that leads to the smallest error is then chosen (line 19), and its more approximated version is then substituted in the main circuit (lines 20-21). The process is then repeated until the error is above the set threshold or all subcircuits are approximated to the highest degree possible.

\section{Experimental Results}
\label{sec:results}
In this section we evaluate our proposed BMF based approximation methodology. Similar to previous work \cite{salsa,aslan14}, we consider a number of arithmetic circuits (adder and multiplier) and a number of application circuits that are amenable for approximate computing such as a multiply-accumulate circuit (MAC), a butterfly network (BUT), a sum of absolute differences (SAD) circuit and finite impulse response (FIR) circuit. Table~\ref{table:apps} summarizes the characteristics of the evaluated applications. Here we also give the number of input and output pins, and the design metrics of the accurate design. To evaluate design area and power consumption, we use Synopsys design compiler with an  industrial 65 nm technology library in typical processing corner.


For all our experiments, as discussed in~\ref{sec:large_cir}, we first decompose each circuit to $k\times m$-cut subcircuits and then perform factorization. In our experiments we chose both $k=10$ and $m=10$. These numbers are simply chosen as they provide a balanced trade-off between truth table complexity and number of subcircuits. We use the modified ASSO algorithm for BMF \cite{Miettinen11,Miettinen14}. Further, for each subcircuit we perform a sweep on the factorization threshold in order to get the best accuracy. In order to evaluate the accuracy on the evaluated applications, we use a Monte Carlo simulation using one million randomly generated input test cases. We define average relative error as,

\vspace{-0.2in}
\begin{equation}
\mbox{Average Relative Error}=\frac{1}{N}\sum_{i=1}^{N} \frac{|R_{i}-R'_{i}|}{R_{i}},
\end{equation}
\vspace{-0.1in}

\noindent and average absolute error as,

\vspace{-0.15in}
\begin{equation}
\mbox{Average Absolute Error}=\frac{1}{N}\sum_{i=1}^{N} |R_{i}-R'_{i}|,
\end{equation}
\vspace{-0.1in}

\noindent
where $N$ is the size of the test case, and $R$ and $R'$ are accurate and approximate results respectively.

Next, in  the first subsection we show the impact of enabling arbitrary QoR functions, when compared to standard $L_2$ metric used in Boolean matrix factorization. In the second subsection, we show the trade-offs and Pareto Frontiers offered by our methodology for our applications.  We also compare the results of  our work to previous work.

\begin{table}[t]
  \footnotesize
  \centering
    \caption{The list of benchmarks evaluated using the proposed NNMF methodology.}
    \vspace{-0.1in}
  \begin{tabular}{c|l|c|r|r|r}
    \hline
      &   &  & \multicolumn{3}{|c}{Accurate Design Metrics} \\
      \cline{4-6}
     Name & Function  & I/O & Area  & Power  & Delay\\
      &  &  & ($um^2$) & ($uW$) & ($ns$)\\
     \hline \hline
     Adder32 & 32-bit Adder   & 64/33 & 320.8 & 81.1 & 3.23 \\\hline
     Mult8 & 8-bit Multiplier & 16/16 & 1731.6	& 263.5 & 2.03 \\\hline
     BUT & Butterfly Structure &  16/18 & 297.4 & 80.6 & 1.79 \\ \hline
     MAC & Multiply and Accumulate &      & & & \\
         & with 32-bit Accumulator &  48/33 & 6013.1 & 470.5 & 2.36 \\\hline
     SAD & sum of absolute  &      & & & \\
         & difference &  48/33 & 1446.5 & 195.1 & 2.43 \\\hline
     FIR & 4-Tap FIR Filter & 64/16 & 8568.0 & 466.3 & 1.56 \\
     \hline \hline
  \end{tabular}
  \vspace{0.1in}
  \label{table:apps}
\end{table}

\begin{figure}[t]
\centering
\scalebox{0.450}{
\epsfig{file=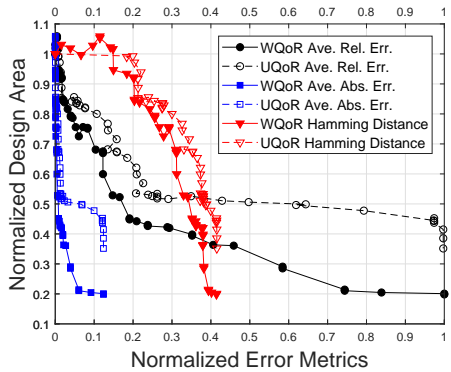}}
\vspace{-0.10in}
\caption{Comparison of the trade-offs offered using the proposed weighted QoR vs. the original factorization algorithm.}
\label{fig:lw_vs_nw}
\end{figure}

\subsection{Evaluation of QoR Impact}
\label{sec:result1}

As described in Section~\ref{sec:methodology}, we modify the Boolean NNMF factorization algorithm, ASSO in this case, to enable weighted cost functions, where a bit error on higher bit indices results in a higher penalty compared to disparities on the lower significance bits.

Figure~\ref{fig:lw_vs_nw} shows the accuracy vs. design area trade-offs offered for the approximate Mult8 design when comparing a factorization algorithm using standard $L_2$ QoR with uniform bit weighting against the proposed weighted QoR. We provide the trends in average relative error, normalized average absolute error, and the normalized Hamming distance. The results obtained from the weighted QoR (WQoR) are shown with solid lines while the dashed lines show the results for the original uniform algorithm (UQoR). 

As shown in the figure, compared to the original algorithm, the weighted scheme provides consistent benefits in accuracy for the same design complexity for all three accuracy metrics. This result confirms the benefit of modifying the BMF algorithm to differentiate among inaccuracies in different indices. Furthermore, this figure highlights the necessity of an algorithm guiding the approximation process in the right direction as suboptimal points are commonly encountered.
Next, we evaluate the trade-offs offered for all of our application circuits using our heuristic design space exploration and compare our results against SALSA~\citep{salsa}.

\subsection{Application Results}
\label{sec:result2}

As previously described, for each application, first the circuit is decomposed into subcircuits with reduced number of inputs and outputs. Then, for each subcircuit and various values of $f$, each subcircuit is approximated and the approximate characteristics are stored. Next, the heuristic proposed in Algorithm~\ref{alg:order} iteratively approximates the subcircuits while assessing the impact on the whole circuit.

\begin{figure*}[t!]
	\begin{center}
		\includegraphics[trim={2.3cm 1cm 0 2cm},scale=0.32]{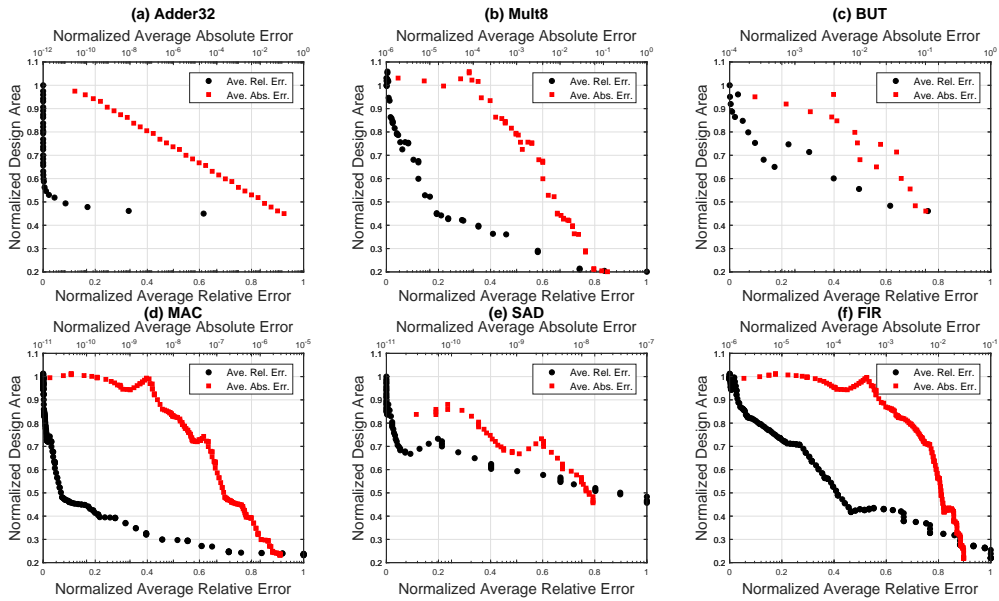}
		\caption{The trade-offs offered for each application. (a) Adder32, (b) Mult8, (c) BUT, (d) MAC, (e) SAD, and (f) FIR.}
		\label{fig:apps_results}
	\end{center}
		\vspace{-0.1in}
\end{figure*}

Figure~\ref{fig:apps_results} shows the trade-offs offered by BLASYS for each of our six benchmarks. In our experiments as the inner workings of accuracy among different blocks is more difficult to model, we simulate the whole circuit while modeling the design metric. More specifically, for design space exploration purposes we assume the design metric, e.g. design area or power, of the large circuit is the sum of design metrics of the $k\times m$-cut subcircuits. For our experiments in order to simplify our design metric model, we use design area as it has less variation compared to power consumption when assembling the subcircuits into the larger circuit. Furthermore, our design area model is only a function of the subcircuits blocks being approximated, while registers and control paths are not considered. We plot the normalized combinational design area utilization as a function of average relative error (black plot and using the bottom x axis) and average normalized absolute error (red plot and using the top x axis). In the case of average absolute error, we normalize the values to the highest output possible. Further, to better show the trend, the average absolute error is plotted in log scale.



As shown in the figure, the proposed methodology enables the designer to choose among a wide range of fine-grain trade-offs.
Intuitively, our design space exploration heuristic aims to find the lowest error possible for a specific degree of approximation where the degree of approximation is incremented by one in each generation. This insight explains the smooth trend of trade-offs for larger circuits while the smaller circuits can change in performance significantly in one iteration. Furthermore, note that while reducing the number of intermediate signals ($f$) generally reduces the complexity of the circuit, there are cases where the number of literals in the logic representation for one output can increase. This phenomenon, explains the temporary increases in design area observed in some of the trends.

The overall runtime of the algorithm is dominated by the accuracy simulation of the intermediate points. Therefore, the runtime is dictated by the Monte Carlo sample size, the threshold set for accuracy, and the tool chain utilized. For example, in our experiments and in the case of the Adder32, the simulation takes about 11 Seconds (using 1 million samples) for each design point, while the BMF algorithm for all the subcircuits takes 0.35 Seconds.

\begin{table}
  \footnotesize
  \centering
    \caption{The hardware characteristics of the approximate testcases for an accuracy threshold of 5\%.}
    \vspace{-0.15in}
  \begin{tabular}{c||r|r|r}
    \hline
          
           &  Area        & Power        & Delay\\
Design     &  Savings (\%) & Savings (\%) & Reduction (\%)\\
\hline \hline
\multirow{ 1}{*}{Adder32} & 44.78 & 63.79 & 12.07 \\ \hline
\multirow{ 1}{*}{Mult8}   & 28.77 & 26.87 & 12.32 \\ \hline
\multirow{ 1}{*}{BUT}     & 7.87  & 11.25 & 2.23  \\ \hline
\multirow{ 1}{*}{MAC}     & 47.55 & 55.58 & 64.41 \\ \hline
\multirow{ 1}{*}{SAD}     & 32.80 & 41.47 & 69.14 \\ \hline
\multirow{ 1}{*}{FIR}     & 19.52 & 22.26 & 12.18 \\ \hline
  \end{tabular}
  \label{table:summary}
\end{table}

Table~\ref{table:summary} summarizes all the design metrics of our 6 testcases and for two accuracy thresholds as synthesized at the end of the design space exploration. As shown in the table, significant reductions in design metrics are possible while insignificant errors are introduced to the circuit.
Based on the application, benefits of approximately 8\%-47\% can be achieved for average relative errors of 5\%.

We also compare our proposed methodology against the previous work SALSA~\citep{salsa}. Table~\ref{table:compare} compares the results obtained using BLASYS against SALSA for given thresholds of 5\% and 25\%. As it can be seen from the table, in all cases, BLASYS delivers significant improvements in design area. We attribute the benefits to BLASYS' ability to approximate multiple outputs, up to $m$ outputs, simultaneously, whereas SALSA approximates each output bit individually. 

\begin{table}[b!]
\vspace{0.05in}
  \footnotesize
  \centering
    \caption{The design area savings at error thresholds 5\% and 25\% for the applications evaluated with comparison to SALSA~\citep{salsa}.}
    \vspace{-0.15in}
  \begin{tabular}{c||r|r|r|r}
    \hline
    & \multicolumn{2}{c|}{Threshold 5\%} & \multicolumn{2}{c}{Threshold 25\%} \\
    & \multicolumn{2}{c|}{Area Savings (\%)} & \multicolumn{2}{c}{Area Savings (\%)} \\
  & \multicolumn{1}{c|}{BLASYS} & \multicolumn{1}{c|}{SALSA} & \multicolumn{1}{c|}{BLASYS}& \multicolumn{1}{c}{SALSA} \\
\hline \hline
Adder32 & 44.9 & 20.5 & 48.2 & 23.2  \\ \cline{2-5}
Mult8 & 28.8 & 1.8 & 63.2 & 8.9 \\ \cline{2-5}
BUT & 7.9 & 5.0 & 26.4 & 24.7  \\ \cline{2-5}
MAC & 47.6 & 1.7 & 65.9 & 8.2  \\ \cline{2-5}
SAD & 32.8 &  3.3 & 38.1 & 15.8 \\ \cline{2-5}
FIR & 19.5 & 3.2 & 34.0 & 15.8 \\ \hline
  \end{tabular}
  \label{table:compare}
\end{table}

\section{Conclusions}
\label{sec:conclusions}
In this paper we proposed a new direction for approximate logic synthesis using boolean matrix factorization. Our proposed methodology, BLASYS, leads to a systematic approach to trade-off accuracy with circuit complexity. To scale our approach into large circuits, we proposed a circuit decomposition heuristic together with a processing order for the subcircuits. Our algorithm results in a very smooth way to trade-off the complexity of entire large circuits with accuracy. We also investigated ways to incorporate different QoR metrics into the circuit factorization algorithm. Our experimental results show solid improvements over state-of-the-art techniques.  

The proposed approach opens many doors for investigation in logic synthesis. Future work include improved techniques for BMF including literal aware approximations, direct incorporation of the QoR metric into the numerical optimization of the factorization algorithm, improved $k\times m$ circuit decomposition, and improved design space heuristics for fewer design point evaluations. Finally, we have integrated our proposed methodology into the open-source YOSYS synthesis framework~\cite{yosys} available on Github \footnote{https://github.com/scale-lab/blasys}.

\noindent\textbf{Acknowledgments:} This work is partially supported by NSF grant 1420864.

\vspace{-0.10in}
\bibliographystyle{ACM-Reference-Format}
\bibliography{refbib2} 

\end{document}